\documentstyle[prl,preprint,epsf,aps]{revtex}
\newcommand{\be}{\begin{equation}}
\newcommand{\ee}{\end{equation}}
\newcommand{\bea}{\begin{eqnarray}}
\newcommand{\eea}{\end{eqnarray}}
\newcommand{\ba}{\begin{array}}
\newcommand{\ea}{\end{array}}

\newcommand{\ep}{\epsilon}
\newcommand{\Ga}{\Gamma}
\newcommand{\ga}{\gamma}
\newcommand{\al}{\alpha}
\newcommand{\ka}{\kappa}
\newcommand{\la}{\lambda}
\newcommand{\de}{\delta}

\newcommand{\om}{\omega}

\newcommand{\br}{{\bf r}}
\newcommand{\Si}{\Sigma}
\newcommand{\ta}{\theta}
\newcommand{\tac}{\theta^{\ast}}
\newcommand{\zc}{z^{\ast}}
\newcommand{\im}{\mbox{Im}}

\newcommand{\cD}{{\cal D}}
\newcommand{\cA}{{\cal A}}

\newcommand{\ub}{\bar{u}}
\newcommand{\vb}{\bar{v}}
\newcommand{\Phib}{\bar{\Phi}}
\newcommand{\psib}{\bar{\psi}}
\newcommand{\chib}{\bar{\chi}}
\newcommand{\bep}{\bar{\epsilon}}

\newcommand{\xic}{\xi^{\ast}}
\newcommand{\w}{{\it w}}

\begin{document}
\draft
\title{Resonant scattering in a strong magnetic field: exact density
of states}
\author{T. V. Shahbazyan and S. E. Ulloa}
\address{Department of Physics and Astronomy,
Condensed Matter and Surface Science Program, Ohio University,
Athens, OH 45701-2979}
\maketitle

\begin{abstract}
We study the structure of 2D electronic states in a strong magnetic
field in the presence of a large number of resonant scatterers. For an
electron in the lowest Landau level, we derive the {\em exact} density
of states by mapping the problem onto a zero-dimensional
field-theoretical model. We demonstrate that the interplay between
resonant and non--resonant scattering leads to a  
{\em non-analytic} energy dependence of the electron Green
function. In particular, for strong resonant scattering the density of
states develops a {\em gap} in a finite energy interval. The shape of
the Landau level is shown to be very sensitive to the distribution of
resonant scatterers. 
\end{abstract}

During recent years there has been a growing interest in the role of
multiple resonant scattering in transport. Most of the studies have been
related to the passage of light through a disordered medium. In
particular, it was shown in a recent experiment\cite{alb91} and subsequent
works\cite{kog92}, that multiple scattering near resonances leads
to a renormalization of the diffusion coefficient up to an order of
magnitude. 

It is natural to expect that resonant scattering would also affect
quite strongly the properties of electrons in disordered
systems. The effective trapping of the electron in resonant states is
expected to suppress diffusion, just as in
optics\cite{alb91,kog92}. This would in turn be evident in the
single--particle density of states (DOS) and localization properties. 

In this paper we study the electronic states of a 2D system in a
strong magnetic field in the presence of a {\em large} number of
resonant scatterers. This choice is motivated in part as 
experimental structures with such geometry became
recently available thanks to remarkable advances in the fabrication of
arrays of ultra--small self-assembled quantum dots\cite{leo93}.  
With typical sizes of less than $20$ nm and very narrow
variations of less than 10\%, an array of such dots with density 
$10^{10}-10^{11}$ cm$^{-2}$ can be produced at some preset distance
from a plane of a high mobility electron gas.\cite{sak95} As the Fermi
energy in the plane approaches the levels of dots, the virtual transitions
between dots and the plane result in multiple resonant
scattering. Such scattering which, in principle, extends through the 
entire system, strongly affects the DOS of a
2D electron.  

The DOS of 2D disordered electronic systems in a quantizing magnetic
field has been extensively studied for the last two 
decades\cite{A,and74,B,D,F,G,I,Q,P,hajdu,T}. The macroscopic
degeneracy of the Landau levels (LL) makes impossible a perturbative
treatment of even weak disorder and calls for non--perturbative
approaches. For high LL, Ando's self--consistent Born
approximation\cite{A} was shown to be asymptotically exact
for short--range disorder\cite{I,P}, while in the case of
long--range disorder the DOS can be obtained within the eikonal
approximation\cite{P}. 
For low LL and uncorrelated disorder, the problem contains no small
parameter and neither of those approximations apply. Nevertheless,
Wegner was able to obtain the exact DOS in a white--noise potential
for the lowest LL, by mapping the problem onto that of the
0D complex $\phi^4$--model\cite{F}. This remarkable result  was
extended to non-Gaussian random potentials by 
Brezin {\em et al.}\cite{G}, and recently to multilayer systems\cite{T}. 

The ``regular'' disorder broadens the LL into a band of width
$\Ga$. At the same time, the resonant scattering leads to a sharp
energy dependence of the DOS near the resonance. The scattering is
enhanced close to the LL center and is suppressed in the
tails. Therefore, the {\em efficiency} of resonant scattering 
is characterized by the ratio $\ga/\Ga$, where $\ga$ is the width of the 
energy spread of resonant states. 

The interplay of the resonant and non--resonant scattering leads to a
rather complex  energy dependence of the DOS.  
Nevertheless, for the lowest LL the problem can be solved {\em exactly} 
[see Eqs.\ (\ref{act4}) and (\ref{dos}) below]. We exploit 
the hidden supersymmetry of the lowest LL\cite{F,G} in order to map
the averaged Green function onto a version of 0D field
theory. The DOS appears to be  {\em non-analytic} as a
function of energy; in particular, it develops a {\em gap} as resonant
scattering becomes strong. 

{\em The model.}---Consider a 2D electron gas separated by a tunneling
barrier from a system of localized states (LS). In addition to LS, a
Gaussian random potential $V(\br)$ with correlator $\langle
V(\br)V(\br')\rangle=w\de(\br-\br')$ is present in the plane.  
We assume that energies of LS are close to the lowest LL and adopt the
tunneling Hamiltonian 

\bea\label{th}
\hat{H}=\sum_{\mu}\ep_{\mu}a^{\dagger}_{\mu}a_{\mu}+
\sum_{i}\ep_{i}c^{\dagger}_{i}c_{i}+
\sum_{\mu,i}(t_{\mu i}a^{\dagger}_{\mu}c_{i}+\mbox{h.c.}),
\eea
where $\ep_{\mu}$, $c^{\dagger}_{\mu}$ and $c_{\mu}$ 
are the eigenenergy, creation and annihilation operators of the 
eigenstate $|\mu\rangle$ of the Hamiltonian $H_0+V(\br)$ ($H_0$
describes a free electron in magnetic field), 
$\ep_{i}$, $c^{\dagger}_{i}$ and 
$c_{i}$ are those of the $i$th LS, and $t_{\mu i}$ 
is a tunneling matrix element. The latter is defined as 
$t_{\mu i}=\int d\br dz
\psi_{\mu}^{\ast}(\br,z)V_{i}(\br,z)\psi_{i}(\br,z)\simeq 
\psi_{\mu}^{\ast}(\br_i,z_i)
\int d\br dzV_{i}(\br,z)\psi_{i}(\br,z)$, where 
$V_{i}(\br,z)$ is the LS potential and $\psi_{i}(\br,z)$ is its wave
function. In the direction normal to the plane, the wave function 
$\psi_{\mu}^{\ast}(\br,z)$ decays as $e^{-\ka z}$, while in the
plane it behaves as an eigenfunction $\psi_{\mu}^{\ast}(\br)$ of the
Hamiltonian $H_0+V(\br)$. For high enough tunneling barrier,
the dependence of $\ka$ on $\mu$ can be neglected\cite{sha94}, so that
$t_{\mu i}\simeq\psi_{\mu}^{\ast}(\br_i)t_i$. 

A formal expression for the Green function of a 2D
electron with energy $\om$, 
$G_{\mu\nu}(\om)=\langle\mu|(\om-\hat{H})^{-1}|\nu\rangle$, 
can be derived by integrating out the LS degrees of freedom. It has
the form $\hat{G}(\om)=(\om-\hat{\ep}-\hat{\Si})^{-1}$, where 
$\hat{\ep}$ is a diagonal matrix with elements $\ep_{\mu}$, 
and self--energy matrix, 

\be\label{self}
\Si_{\mu\nu}(\om)=\sum_i{t_{\mu i}t_{i \nu}\over \om-\ep_i}=
\sum_i{t_{i}^2\psi_{\mu}^{\ast}(\br_i)\psi_{\nu}(\br_i)
\over \om-\ep_i}.
\ee
comes from scattering of the electron by LS. In such a form,
however, the Green function is hard to analyze. Instead, it is
convenient to work with an effective {\em in--plane} Hamiltonian,
$H_{\rm eff}$, for the electron with energy $\om$. 
Recasting $\Si_{\mu\nu}$ in coordinate 
representation, we obtain
$H_{\rm eff}(\om)=H_0+V(\br)+U(\om,\br)$, where the last term,

\be\label{vres}
U(\om,\br)=\sum_i{t_i^2\over \om-\ep_i}\de(\br_i-\br),
\ee
describes the resonant scattering of electron by the LS. The
potential (\ref{vres}) resembles that of point--like 
scatterers. The crucial difference, however, is that here  
scattering strength depends on the proximity of the
electron energy to the LS levels. It is important to notice that
$U(\om,\br)$ changes from repulsive to attractive as the
electron energy passes through the resonance. Since positions of LS
are random with uniform density $n_{_{LS}}$, the distribution function
of $U$ is Poissonian.

In the following, we assume that the tunneling barrier is high enough,
and neglect the difference in $t_i$ for different LS, setting 
$t_i=t$ in the rest of the paper. Strong magnetic field 
implies that scattering keeps electron in the lowest LL. While for the
white--noise potential this condition is standard, it is  
more restrictive for the resonant scattering. It should be noted,
however, that the latter is effectively reduced by the energy spread of LS. 

The calculation of the DOS,
$g(\om)=-\pi^{-1}\im\overline{G(\br,\br)}$, requires averaging of the
Green function, 
$G(\br,\br)=\langle\br|(\om_{+}-H_{\rm eff})^{-1}|\br\rangle$ 
(with $\om_{+}=\om+i0$) over {\em both} random potentials
$V(\br)$ and $U(\om)$. Below, we derive this DOS exactly by
using the approach of Ref.~\cite{G}.

{\em Derivation of DOS.}---The Green function is presented 
as a bosonic functional integral 
$G(\br,\br)=-iZ^{-1}\int\cD\psi\cD\psib 
e^{iS}\psi(\br)\psib (\br)$ with the action
$S[\psib ,\psi]=\int d\br 
\psib (\br)[\om_{+}-H_{\rm eff}(\om)]\psi(\br)$.
After writing the normalization factor as a fermionic
integral $Z^{-1}=\int\cD\chi\cD\chib e^{iS}$ with the
same action $S[\chib ,\chi]$, both $\psi$ and
$\chi$ are projected on the lowest LL subspace
as $(\om-H_{0})\psi=\om\psi$ 
(measuring all energies from the lowest LL). In the symmetric gauge,
this projection is achieved with 
$\psi=(2\pi l^2)^{-1/2}e^{-|z|^2/4l^2}u(z)$ and 
$\chi=(2\pi l^2)^{-1/2}e^{-|z|^2/4l^2}v(z)$, where the bosonic field
$u(z)$ and the fermionic field $v(z)$ are analytic functions of the
complex coordinate $z=x+iy$ ($l$ is the magnetic
length). The  Green function then takes the form  
$G(\br,\br)=-i(2\pi l^2)^{-1}e^{-|z|^2/2l^2}
\langle u(z)\ub (\zc )\rangle$, where $\langle\cdots\rangle$ denotes
a functional integral over $u(z)$ and $v(z)$ with the action

\be\label{act2}
S=\int{d^{2}z\over 2\pi l^2}e^{-|z|^2/2l^2}(\ub u+\vb v)
[\om_{+}-V-U(\om)].
\ee
As a next step, one introduces Grassman coordinates $\ta$ and $\tac$,
normalized as $\int d^2zd^2\ta e^{-|z|^2-\ta\tac}=1$, and defines 
analytic ``superfields''  $\Phi(z,\ta)=u(z)+\ta v(z)/\sqrt{2}l$ and 
$\Phib (\zc ,\tac)=\ub (\zc )+\tac \vb (\zc )/\sqrt{2}l$, taking
values in the ``superspace'' $\xi=(z,\ta)$. Using 
$\langle u\rangle=\langle v\rangle=0$ and 
$\langle u\ub\rangle=\langle v\vb\rangle$, the Green function can be
presented as

\bea\label{green}
G=-i{e^{-\xi\xic/2l^2}\over 2\pi l^2}\int\cD\Phi \cD\Phib 
e^{iS}\Phi(\xi)\Phib (\xic), 
\eea
where $\xi\xic\equiv |z|^2+\ta\tac$ and 
$S[\Phib,\Phi]$ is obtained from (\ref{act2}) by substituting 
$\ub u+\vb v=2\pi l^2\int d^2\ta e^{-\ta\tac/2l^2}
\Phib (\xic)\Phi(\xi)$.

We now perform the ensemble averaging over $V$ and $U$. The
Gaussian averaging of $\exp i\int V Q d^2 z$, where
$Q=\int d^2\ta e^{-\xi\xic/2l^2}\Phib(\xic)\Phi(\xi)$,
gives $\exp\left[-(\w/2)\int Q^2 d^2z\right]$, while the averaging
of $\exp i\int U Q d^2 z$ with a Poissonian distribution function
yields\cite{fri75} 

\be\label{aver}
\exp \biggl\{-n_{ls}\int \biggl[1-
\biggl\langle \exp\biggl(-{it^2 Q\over \om-\ep}\biggr)
\biggr\rangle_{\ep}\biggr]d^2z\biggr\}, 
\ee
where $\langle\cdots\rangle_{\ep}$ denotes energy averaging.
As a result, one obtains the following effective action

\bea\label{act3}
iS
&&
[\Phi,\Phib]=i\om_{+}\int d^2\xi\, \al
-{\Ga^2\over 2}\int {d^{2}z\over 2\pi l^2}
\left(2\pi l^2\int d^2\ta\, \al \right)^2
\nonumber\\
&&
-\nu\int {d^{2}z\over 2\pi l^2}\left\{1-\left\langle
\exp\left[-i\la\, 2\pi l^2\int d^2\ta\,\al 
\right]\right\rangle_{\ep}\right\},
\eea
where $\al(\xi,\xic)=e^{-\xi\xic/2l^2}\Phib(\xic)\Phi(\xi)$.
Here $\Ga=(\w/2\pi l^2)^{1/2}$ is Wegner's
width of lowest LL (in the absence of resonant scattering),
$\nu=2\pi l^2n_{_{LS}}$ is the ``filling factor'' of LS, and we
denoted $\la=\de^2/(\om-\ep)$, where $\de=t/(2\pi l^2)^{1/2}$ 
characterizes the tunneling, 

The action (\ref{act3}) possesses a supersymmetry, 
characteristic for the lowest LL\cite{F,G}. Being evident for the
first term, this symmetry between $z$ and $\ta$ can be made explicit
for the second and third terms 
also by making use of the identity\cite{G}
$n\left(2\pi l^2 \int d^2\ta e^{-\ta\tac/2l^2} \Phib \Phi \right)^n
= 2\pi l^2 \int d^2\ta e^{-n\ta\tac/2l^2} (\Phib\Phi)^n$, which allows
one to replace any functional of the form 
$\int d^2z f\left(2\pi l^2\int d^2\ta \al\right)$ with
$2\pi l^2\int d^2\xi h\left(\al\right)$,
where $\partial h(x)/\partial x =f(x)/x$. As a  result one obtains a 
manifestly supersymmetric action $S=\int d^2\xi\cA(\al)$, where

\bea\label{act4}
i\cA(\al)=i\om_{+}
&&
\al-{\Ga^2\al^2\over 4}
\nonumber\\
&&
-\nu\int_0^{\al}{d\beta\over\beta}\biggl[1-\biggl\langle
\exp\biggl(-{i\de^2 \beta\over \om-\ep}\biggr)
\biggr\rangle_{\ep}\biggr].
\eea
The supersymmetry leads, in turn, to the exact cancelation of contributions
from $z$ and $\ta$ spatial integrals into each diagram, so that the
entire perturbation series can be  
generated in the 0D field theory with the {\em same} action\cite{F,G}.
The Green function is then given by the ratio of two ordinary integrals, 
$G(\om)=-i(2\pi l^2)^{-1}Z_0^{-1}\int d^2\phi e^{i\cA}\phi\phi^{\ast}$,
where $Z_0=\int d^2\phi e^{i\cA}$ with $\cA(\phi\phi^{\ast})$ from
(\ref{act4}). From this Green function, the DOS is obtained as

\bea\label{dos}
g(\om)={1\over 2\pi^2 l^2}\im{\partial\over\partial \om_{+}}
\ln\int_0^{\infty}d\al e^{i\cA(\al)},
\eea
where the derivative applies only to the
first term of (\ref{act4}). 

{\em Examples.}---The energy averaging in (\ref{act4}) can be
performed analytically for an arbitrary distribution of LS levels,
$f_{\ga}(\ep-\bep)$, where $\bep$ is average energy and $\ga$ is the
width. The result reads 

\bea\label{exp}
i\cA(\al)
&&
=i\om_{+}\al-
{\Ga^2\al^2\over 4}
\nonumber\\
&&
-\nu\int_0^{\infty}{dx\over x}\tilde{f}_{\ga}(x) e^{i(\om-\bep)x}
\left[1-J_0\left(2\de\sqrt{x\al}\right)\right],
\eea
where $\tilde{f}_{\ga}(x)$ is Fourier transform of $f_{\ga}(\ep)$ and
$J_0$ is the Bessel function. Numerical results for DOS with Gaussian
distribution, $\tilde{f}_{\ga}(x)=e^{-\ga x^2/2}$, are
presented in Fig.~1.  

Consider first the case of a strong in--plane disorder, 
$\Ga/\de\gg 1$. For a not 
very small $\ga$, so that $\de^2/\ga\Ga\ll 1$, the Bessel function in
(\ref{exp}) can be expanded to first order, yielding
$G(\om)=G_W(\om-\Si)$, where $G_W(\om)$ is Wegner's Green function
(that is with $\nu=0$) and   
$\Si(\om)=-i\nu\de^2\int_0^{\infty}dx e^{-\ga^2x^2/2+i(\om-\bep)x}$
is the first--order self-energy due to the resonant scattering. If the
resonant level is close to the LL center, $\om\sim\bep\ll \Ga$, the
first-order correction to the DOS reads 

\bea\label{dg}
{\de g(\om)\over g_W(0)}=-{\pi-2\over\sqrt{2}}{\nu\de^2\over\ga\Ga}
\exp \left[-{(\om-\bep)^2\over 2\ga^2}\right],
\eea
where $g_W(\om)$ is Wegner's DOS.

Resonant scattering in this case manifests itself as a minimum of 
width $\ga$ on top of the wider peak of width $\Ga$.  
The evolution of the DOS with increasing $\de/\ga$ is shown in
Fig.~1(a). The effect is strongest for $\de/\ga \gg 1$, 
however splitting remains considerable even for
$\ga/\de\simeq 1$. For $\de/\ga \ll 1$ the DOS is basically
unaffected by resonant scattering and reduces to Wegner's form 
$g_W(\om)$.

With increasing scattering $\de/\Ga$, the shape of the DOS undergoes
drastical transformation [see Fig.1(b)]. For a strong scattering, the
DOS develops a {\em gap} in the energy interval $\om(\om-\bep)<0$. The
existence of the gap can be traced directly to Eq.~(\ref{act4}) 
(with vanishing $\ga/\de$ and $\Ga/\de$). In this energy interval the
integration path in the $\al$--integral in (\ref{dos}) can be rotated 
by $e^{-i\pi {\rm sgn}(\om-\bep)/2}$, resulting in a purely real $i\cA$. 
The origin of the gap is the following. If the  ``regular''
disorder is weak (small $\Ga$), the LL broadening comes from the
resonant scattering alone. Then the scattering potential
(\ref{vres}) appears to be attractive for $\om<\bep$, pulling the
electronic states from the LL center to the {\em left}, while for
$\om>\bep$ the potential is repulsive, pushing the states to the {\em
right}. Note that for a low density of scatterers, $\nu<1$, a fraction
$1-\nu$ of states in the plane remains unaffected. Such
``condensation of states'' was known also for the case of repulsive
point--like scatterers with a constant scattering
strength\cite{and74,B,G,I}. In fact, the analogy extends also to the
intricate structure of the DOS away from the gap. In particular, the
smaller peaks correspond to singularities in $g(\om)$ at
integer values of $\om(\om-\bep)/\de^2$\cite{G}. The behavior of
$g(\om)$ near the gap edges is different for $\om\rightarrow 0$ and 
$\om\rightarrow \bep$: one can show that in the former case the DOS
exhibits a discontinuity,   
$g(\om)\propto (1-\nu)\de(\om)+\mbox{const}/|\om|^{\nu}$, while near
the resonance it vanishes as $(\om-\bep)^{1-\nu}$. With increasing
$\ga$, the gap and the smaller peaks are washed out; however the peak at 
$\om=0$ persists throughout [see Fig.\ 1(c)].

In conclusion, although our derivation was restricted to the lowest LL,
we believe that our results are more general and valid for
higher LL also. Indeed, the gap in the DOS for small disorder is
apparently a result of the LL degeneracy. Therefore, the above
argument, related to the change in the sign of the potential
(\ref{vres}), should hold for arbitrary LL. Note that the 
``condensation of states'' also occurs for all LL numbers\cite{I}. Thus, 
we expect that the gap in the DOS will persist, although the precise
behavior of $g(\om)$ near the gap edges could be different. 
Concerning the sharp minimum in the DOS in the absence of the LL
degeneracy [see Fig. 1(a)], it seems that this is a rather general
feature. In fact, in the absence of magnetic field, analogous behavior 
has been known in the 3D case for identical scatterers \cite{iva,jau83}.

A possible experimental realization of the multiple resonant
scattering could be a system of self-assembled quantum dots
separated from a 2D electron gas by a tunable tunneling
barrier \cite{sak95}. Due to the narrow distribution of dots'
sizes, the spread in their energy levels, $\ga$, does not exceed 
10 meV\cite{leo93}. For considerable effect of the resonant
scattering, one must have  
$\de^2/\ga\Ga\sim 1$. For a typical LL width $\Ga\sim 1$ meV, this
condition implies that the parameter $\de$ should be about
several meV, which would be reasonable to achieve. Moreover, for 
$\de/\Ga\gtrsim 1$, an even weaker condition, the tunneling
would be relatively strong and the effect of the resonant scattering
would be significant.  It was observed in Ref.\ \cite{sak95} that the
mobility of the 2D gas (at zero magnetic field) dropped by two orders of
magnitude when the thickness of tunneling barrier between the dots and
the plane was reduced. Although, we cannot give quantitative estimate
for the zero--field case, this is certainly in qualitative agreement
with our results.  We hope that our results would further motivate
experiments in magnetic fields.

Finally, we have disregarded the possible charging
effects and assumed that the transitions occur between the plane and
unoccupied dots. Certainly, as the Fermi energy approaches
$\bep$ some of the dots will become singly occupied. Once occupied,
such dots would have much higher energies and would  not participate
in the resonant scattering, reducing the effective density
$n_{_{LS}}$, apart from producing (uniform) Coulomb shifts in the
energies of unoccupied dots. 

This work was supported in part by the US Department of Energy Grant
No. DE-FG02-91ER45334.


\begin{figure}
\caption{
(a) DOS [in units of $g_1=(2\pi l^2)^{-1}\Ga^{-1}$] for
strong in-plane disorder, $\de/\Ga=0.3$, with $\bep=0$ and $\nu=1.5$,
is shown for different $\ga/\de=0.1$ (solid line), 0.5 (dotted), 
1.0 (dashed), 2.0 (long--dashed), and 10.0 (dot--dashed). 
(b) DOS [in units of $g_2=(2\pi l^2)^{-1}\de^{-1}$] for strong
tunneling, $\de/\ga=10.0$, with $\bep=\de$ and $\nu=0.8$, is shown
for $\Ga/\de=0.1$, (solid line), 0.2 (dotted),
0.3 (dashed), 0.5 (long--dashed), and
1.0 (dot--dashed).
(c) The DOS for weak in-plane disorder, $\Ga/\de=0.1$, with 
$\bep=\de$ and $\nu=0.8$, is shown for $\ga/\Ga=1.0$, (solid line), 
3.0 (dotted), 5.0 (dashed), and 10.0 (long--dashed).}
\vspace{-0.26cm}
\label{fig:1}
\end{figure}


\clearpage

\begin{figure}[htb]
\epsfxsize=6.5in
\epsfbox{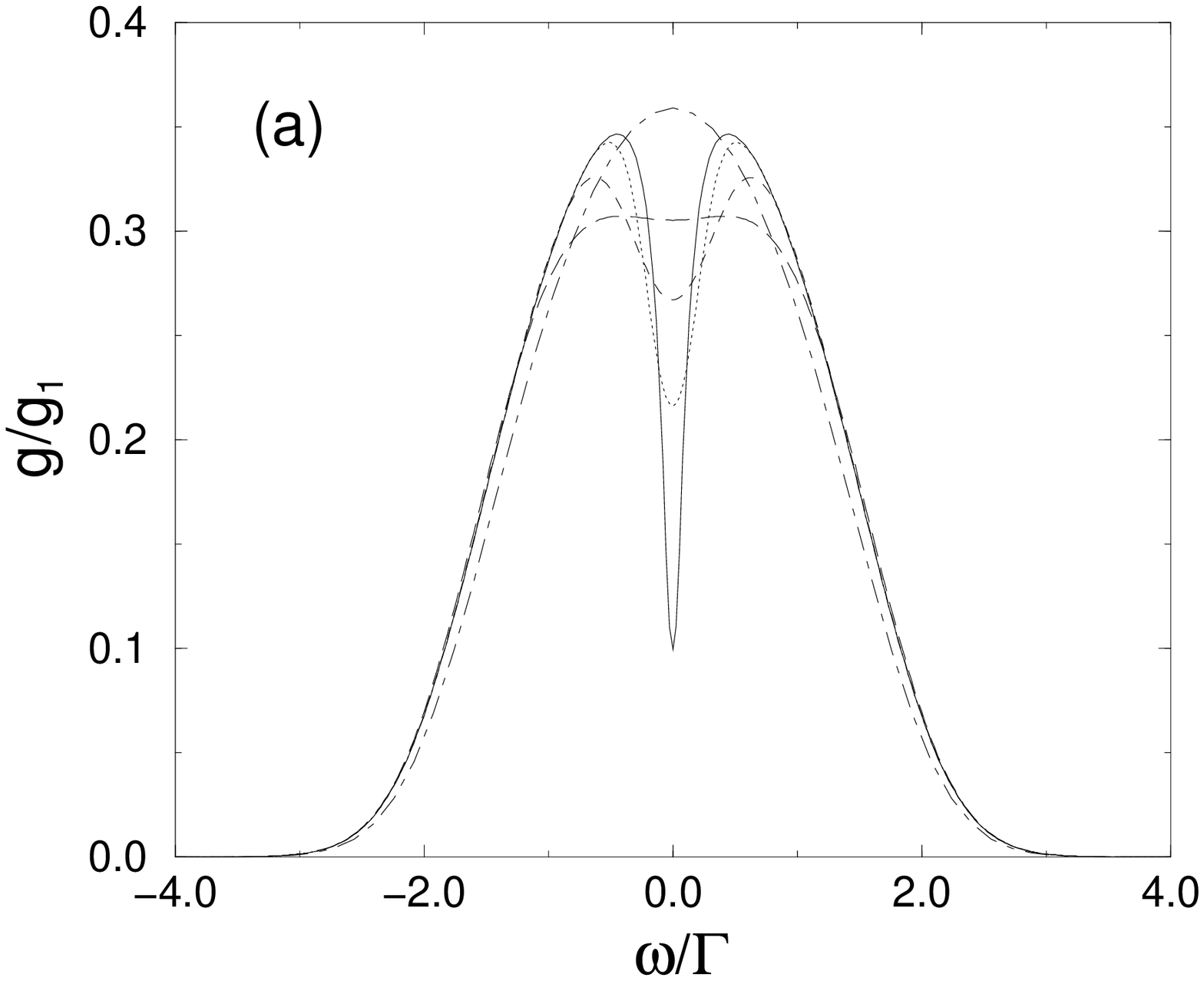}
\end{figure}

\clearpage

\begin{figure}[htb]
\epsfxsize=6.5in
\epsfbox{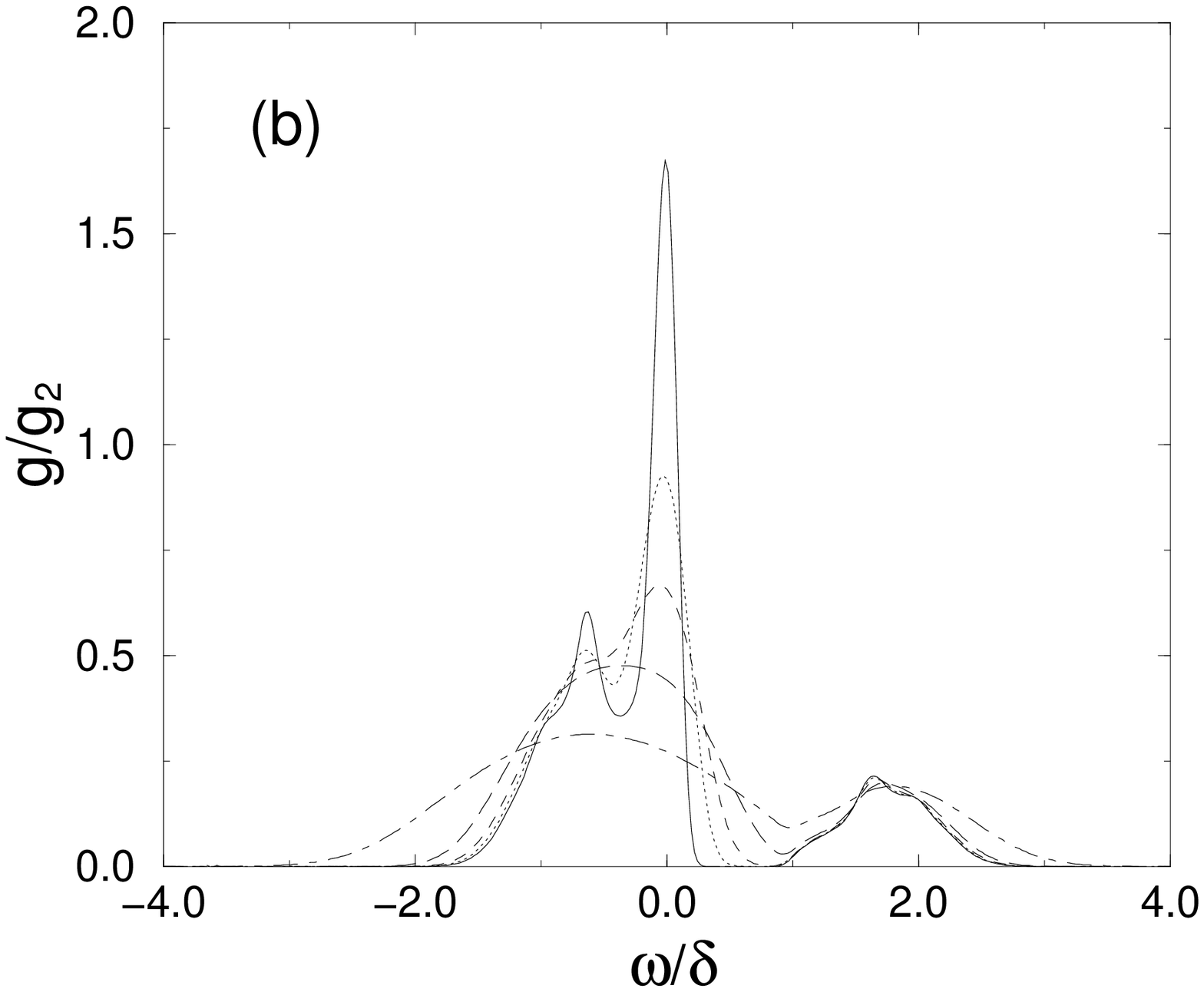}
\end{figure}

\clearpage

\begin{figure}[htb]
\epsfxsize=6.5in
\epsfbox{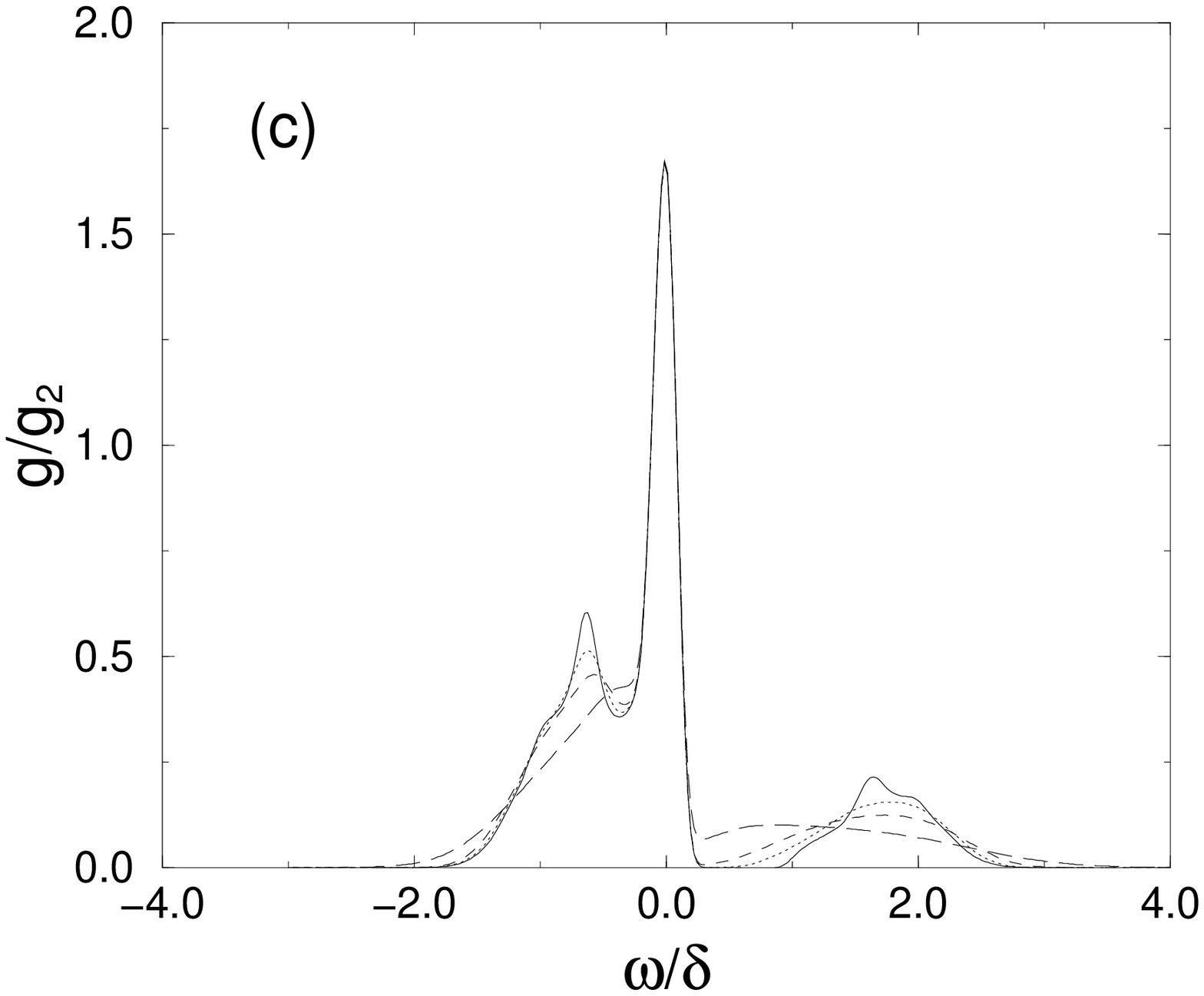}
\end{figure}


\begin{references}

\bibitem{alb91}M.\ P.\ van\ Albada, {\em et al.},
Phys.\ Rev.\ Lett.\ {\bf 66}, 3132 (1991);
B.\ A.\ van Tiggelen, {\em et al.},
Phys.\ Rev.\ B\ {\bf 45}, 12 233 (1992).

\bibitem{kog92}E. Kogan and M. Kaveh,
Phys. Rev. B {\bf 46}, 10 636 (1992);
G. Cwilich and Y. Fu,
{\em ibid.}\ {\bf 46}, 12 015 (1992);
Yu. N. Barabanenkov and V. Ozrin,
Phys.\ Rev.\ Lett.\ {\bf 69}, 1364 (1992);
B.\ A.\ van Tiggelen, {\em et al.},
{\em ibid.}\ {\bf 71}, 1284 (1993);
K. Busch and C. M. Sokoulis,
{\em ibid.}\ {\bf 75}, 3442 (1995).

\bibitem{leo93}D. Leonard, {\em et al.}, 
Appl. Phys. Lett. {\bf 63}, 3203 (1993);
H. Drexler, {\em et al.},
Phys. Rev. Lett. {\bf 73}, 2252 (1994);
J.-Y. Marzin, {\em et al.},
{\em ibid.}\ {\bf 73}, 716 (1994);
M. Grundmann, {\em et al.},
{\em ibid.}\ {\bf 74}, 4043 (1995);
S. Fafard, {\em et al.},
Phys. Rev. B {\bf 50}, 8086 (1994);
P. D. Wang, {\em et al.},
{\em ibid.}\ {\bf 53}, 16 458 (1996).

\bibitem{sak95}H. Sakaki, {\em et al.}
Appl. Phys. Lett. {\bf 67}, 3444 (1995).

\bibitem{A} T. Ando and Y. Uemura,
J.\ Phys.\ Soc.\ Jpn.\ {\bf 36}, 959 (1974).

\bibitem{and74} T. Ando,
J.\ Phys.\ Soc.\ Jpn.\ {\bf 36}, 1521 (1974);
{\bf 37}, 622 (1974); {\bf 37}, 1233 (1974).

\bibitem{B} \'{E}. M. Baskin, L. N. Magarill and M. V. \'{E}ntin,
Zh.\ Eksp.\ Teor.\ Fiz.\ {\bf 75}, 723 (1978) 
[Sov.\ Phys.\ JETP\ {\bf 48}, 365 (1978)].

\bibitem{D} L. B. Ioffe and A. I. Larkin,
Zh.\ Eksp.\ Teor.\ Fiz.\ {\bf 81}, 1048 (1981) 
[Sov.\ Phys.\ JETP\ {\bf 54}, 556 (1981)];
I. Affleck, J.\ Phys.\ C\ {\bf 16}, 5839 (1983);
{\bf 17}, 2323 (1984).

\bibitem{F} F. Wegner, Z.\ Phys.\ B\ {\bf 51}, 279 (1983).

\bibitem{G} E. Br\'{e}zin, D. Gross, and C. Itzykson,
Nucl.\ Phys.\ {\bf B235}, 24 (1984).

\bibitem{I} K. A. Benedict and J. T. Chalker, 
J.\ Phys.\ C\ {\bf 18}, 3981 (1985); {\bf 19}, 3587 (1986);
K. A. Benedict, Nucl.\ Phys.\ {\bf B280}, 549 (1987).

\bibitem{Q}
S. A. Gredeskul, Y. Avishai, and M. Ya Azbel', 
Europhys. Lett. {\bf 21}, 489 (1993);
Y. Avishai, M. Ya. Azbel', and  S. A. Gredeskul, 
Phys.\ Rev.\ B\ {\bf 48},\ 17280\ (1993);
Zusman,\ Y. Avishai,\ and S. A. Gredeskul,
{\em ibid.}\ {\bf 48},\ 17922\ (1993).

\bibitem{P}M. E. Raikh and T. V. Shahbazyan,
Phys.\ Rev.\ B\ {\bf 47}, 1522 (1993).

\bibitem{hajdu}M.\ Jan{\ss}en\ {\em et al.}
{\em Introduction to the Theory of the Integer Quantum Hall
Effect}, (VCH, Weinmheim, 1994).

\bibitem{T}T. V. Shahbazyan and M. E. Raikh,
Phys.\ Rev.\ Lett.\ {\bf 77}, 5106 (1996).

\bibitem{sha94}T. V. Shahbazyan and M. E. Raikh, 
Phys. Rev. B {\bf 49}, 17 123 (1994).

\bibitem{fri75}R. Friedberg and J. M. Luttinger,
Phys. Rev. B {\bf 12}, 4460 (1975).

\bibitem{iva}M. A. Ivanov and Yu. G. Pogorelov
Zh.\ Eksp.\ Teor.\ Fiz.\ {\bf 76}, 1010 (1979)
[Sov.\ Phys.\ JETP\ {\bf 49}, 510 (1979)].

\bibitem{jau83}A. P. Jauho and J. W. Wilkins
Phys. Rev. B {\bf 28}, 4628 (1983).


\end{references}
\end{document}